\documentclass[aps,twocolumn,amsmath,amssymb,showpacs,showkeys,superscriptaddress,notitlepage,longbibliography]{revtex4-1}
\usepackage[colorlinks=true,linkcolor=blue,anchorcolor=red,citecolor=blue, urlcolor=blue]{hyperref}
\usepackage{bm}
\usepackage{graphicx}
\usepackage{color}
\usepackage{makecell}
\usepackage{multirow}

\begin{document}
\title{Recent progress on Majorana in semiconductor-superconductor heterostructures --- engineering and detection}
\author{Zhan Cao}
\affiliation{Beijing Academy of Quantum Information Sciences, Beijing 100193, China}

\author{Shumeng Chen}
\affiliation{State Key Laboratory of Low Dimensional Quantum Physics, Department of Physics, Tsinghua University, Beijing, 100084, China}
\affiliation{Frontier Science Center for Quantum Information, Beijing 100084, China}

\author{Gu Zhang}
\affiliation{Beijing Academy of Quantum Information Sciences, Beijing 100193, China}

\author{Dong E. Liu}
\email{dongeliu@mail.tsinghua.edu.cn}
\affiliation{State Key Laboratory of Low Dimensional Quantum Physics, Department of Physics, Tsinghua University, Beijing, 100084, China}
\affiliation{Beijing Academy of Quantum Information Sciences, Beijing 100193, China}
\affiliation{Frontier Science Center for Quantum Information, Beijing 100084, China}

\begin{abstract}
Majorana zero modes (MZMs) are exotic excitations (in condensed matter systems) of fundamental scientific interest and hold great promise as the basis of fault-tolerant topological quantum computation.
Though MZMs have been predicted in many platforms, their existence is still under debate. In this paper, we review the recent progress in engineering and detecting MZMs in semiconductor-superconductor heterostructures. We also briefly review the protocols for implementing topological quantum computation by hybrid semiconductor-superconductor nanowires.
\end{abstract}



\maketitle

\section{Introduction}\label{int}
Majorana zero mode (MZM), also known as Majorana bound state, is an exotic zero-energy excitation with equal electron and hole weights~\cite{read2000paired,kitaev2001unpaired}, in analog with Majorana fermion~\cite{majorana1937teoria} whose antiparticle is itself. So far, no Majorana fermion as elementary particles has been found in nature, however, MZMs are predicted to exist in topological superconductors (SCs), please refer to reviews~\cite{alicea2012new,leijnse2012introduction,stanescu2013Majorana,beenakker2013search,beenakker2015random,elliott2015colloquium,sato2016majorana,aguado2017Majorana,sato2017topological,lutchyn2018majorana,beenakker2020search,flensberg2021engineered,fu2021experimental,marra2022bMajorana}.
In recent years, MZMs have attracted broad interdisciplinary interest due to their unique exchange statistics and potential for realizing fault-tolerant topological quantum computation~\cite{kitaev2003fault,nayak2008non,sarma2015majorana}. 

MZMs are predicted to exist at the interfaces of materials with different topology,
or in the vortices of a 2D topological SC (see, e.g.,  Ref.~\onlinecite{alicea2012new} for a pedagogical review). 
Motivated by its fundamental scientific interest and promising application in quantum computation, MZMs and their host, i.e., topological SCs, have received considerable experimental and theoretical interest in the last decade.
To date, topological SC candidates can be classified into intrinsic and artificial ones. Since 2010, abundant intrinsic topological SC candidates have been experimentally reported, including, e.g., putative spinless chiral $p$-wave SC (Sr$_2$RuO$_4$~\cite{maeno2001intriguing,sarma2006proposal}), metal-doped topological insulators (Cu$_x$Bi$_2$Se$_3$~\cite{wray2010observation,sasaki2011topological}, Sr$_x$Bi$_2$Se$_3$~\cite{liu2015superconductivity}, Nb$_x$Bi$_2$Se$_3$~\cite{asaba2017rotational}), iron-based SCs (FeTe$_{0.55}$Se$_{0.45}$~\cite{wang2018evidence}, Li(Fe,Co)As~\cite{zhang2019multiple}, (Li,Fe)OHFeSe~\cite{liu2018robust}), transition metal dichalcogenides (2M-WS$_2$~\cite{yuan2019evidence}), non-centrosymmetric SCs (PbTaSe$_2$~\cite{bian2016topological}), and centrosymmetric SCs ($\beta$-Bi$_2$Pd~\cite{lv2017experimental}).
Meanwhile, a variety of artificial topological SC candidates have been experimentally studied, including, e.g., the topological insulator-$s$-wave SC heterostructure (Bi$_2$Te$_3$-NbSe$_2$~\cite{xu2015experimental,sun2016majorana}, HgTe-Nb~\cite{wiedenmann20164pi}, HgTe-Al~\cite{bocquillon2017gapless,deacon2017josephson}), ferromagnetic atomic chain on the surface of an $s$-wave SC (Fe-Pb~\cite{nadj2014observation}), semiconductor (SM) nanowire-$s$-wave SC heterostructure (see Sec.~\ref{Sec II} for a review), and ferromagnetic insulator-metal-$s$-wave SC heterostructure (EuS-Au-V \cite{manna2020signature}).

Despite the ongoing search for MZMs in the last decade,
the question of whether we are witnessing proven evidence of MZMs is still being debated. Currently, SM-SC heterostructures have become the most studied topological SC candidates, as (i) they require only conventional materials and are compatible with the well-developed SM growth techniques, and (ii) the roadmap for implementing quantum computation with these SM techniques has been established~\cite{aasen2016milestones,plugge2017majorana,karzig2017scalable}.
In this paper, we review recent theoretical and experimental progress on engineering and detecting MZMs in SM-SC heterostructures.
We also briefly review the related protocols for implementing quantum computation.

This paper is organized as follows. We review the available SM-SC heterostructures for engineering MZMs in Sec.~\ref{Sec II} and the experimental protocols that have been performed to detect MZMs in Sec.~\ref{Sec III}. In Sec.~\ref{Sec IV}, we review the recent progress of material choice and growth of SM-SC heterostructures. In Sec.~\ref{Sec V}, we provide a preliminary introduction to topological quantum computation protocols implemented by hybrid SM-SC nanowires. Finally, we give in Sec.~\ref{Sec VI} a brief summary and the prospective challenges in this field.

\section{Recent progress on engineering MZMs in SM-SC heterostructures}\label{Sec II}
In 2010, R. M. Lutchyn \textit{et al.}~\cite{lutchyn2010majorana} and Y. Oreg \textit{et al.}~\cite{oreg2010helical} independently proposed to engineer MZMs in a quasi-1D SM nanowire proximitized to an $s$-wave SC, under the influence of a strong Rashba spin-orbit coupling and a magnetic field induced Zeeman splitting.
This Majorana-nanowire proposal is rather attractive as it requires only conventional SM, e.g., InAs or InSb, and $s$-wave SC, e.g., Al, to equivalently realize the MZM-hosted Kitaev chain proposed by the pioneering work Ref.~\cite{kitaev2001unpaired}, where instead an unconventional $p$-wave SC is demanded.

\begin{figure}[t!]
\centering
\includegraphics[width=\columnwidth]{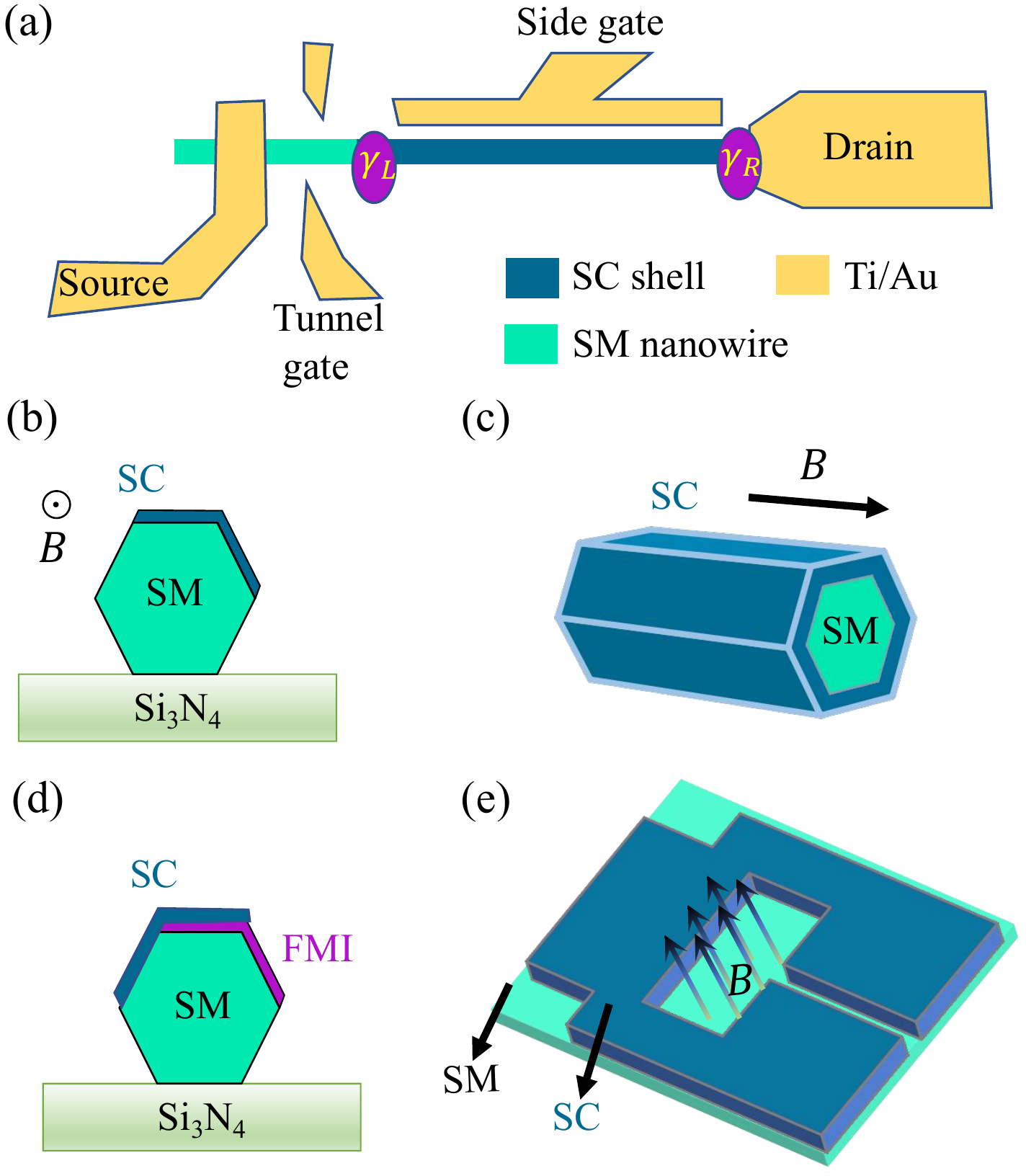}
\caption{(a) Top view of the schematic of the tunneling spectroscopy measurement of a hybrid SM-SC nanowire that is predicted to host a pair of MZMs at the opposite wire ends.
The tunnel gate controls the tunnel barrier between the source lead and the hybrid nanowire, while the side gate tunes the electron density in the SM nanowire.
The hybrid nanowire can be either the essential quasi-1D ones shown by (b)--(d) or the one etched on a 2D SM-SC heterojunction.
(b)--(e)  Cross section of a hybrid SM-SC nanowire on a dielectric material, in which a thin SC shell is epitaxially grown on pristine facets of the SM nanowire.
Note that the nanowire is partially covered by the SC shell such that the chemical potential of the nanowire can be tuned by voltage gates. A magnetic field $B$ in parallel to the nanowire is necessary to induce a Zeeman splitting in the SM.
(c) Side view of a full-shell hybrid SM-SC nanowire.
Instead of a Zeeman splitting induced by a magnetic field, the magnetic flux threading through the cross-section of the SM core creates fluxoids in the superconducting order parameter of the SC shell and drives the system into the topological superconducting phase. (d) Cross section of an SM-FMI-SC nanowire on a dielectric material. No magnetic field is needed, while the magnetic proximity effects at the SM-FMI and SC-FMI interfaces are both essential for achieving topological superconductivity. (e) Planar JJ formed in a 2D SM covered by an SC shell. A pair of MZMs are predicted to localize at two ends of the stripe region uncovered by the SC shell. By threading a magnetic flux through the nearby superconducting loop, the phase difference between the superconducting regions on the two sides of the stripe can be tuned to drive the system into the topological superconducting phase with a tiny magnetic field.}\label{Fig:device}
\end{figure}

In 2012, the device based on the Majorana-nanowire model was firstly fabricated~\cite{mourik2012signatures} by placing part of an InSb nanowire on top of a bulk SC NbTiN. Multiple metallic gates are placed next to the wire to control both the chemical potential and the tunneling through a potential barrier. An external magnetic field is also applied to generate the required Zeeman splitting. Several transport features were observed in the tunneling spectroscopy~\cite{mourik2012signatures}, including (i) the appearance of a zero-bias conductance peak (ZBCP) under a parallel magnetic field ranging from 0.07 to 1 T; (ii) the ZBCP remains when changing the applied voltages of different gates over large ranges; (iii) the strong dependence of the zero-bias conductance on the direction of the applied magnetic field; and (iv) the missing of ZBCP after replacing the SC by a normal metal. These observations coincide with the MZM signatures predicted by the Majorana-nanowire model. Afterwards, similar results were reported in a few experiments using InSb or InAs nanowires~\cite{deng2012anomalous,das2012zero,finck2013anomalous,churchill2013superconductor}. In these experiments, substantially manifest subgap conductance coexists with the ZBCP. This phenomenon, known as soft-gap, is detrimental to topological quantum computation as it degrades the topological protection. Disorder impacts at the SM-SC interface are believed to cause soft gap features~\cite{takei2013soft}. In 2015, with the technical advances, these sub-gap states are dramatically suppressed in epitaxially grown hybrid SM-SC nanowires~\cite{chang2015hard,krogstrup2015epitaxy}, where a thin SC shell is directly grown on the pristine facets of the SM nanowire [Fig.~\ref{Fig:device}(a)] to greatly improve the cleanness of the SM-SC interface. Note that the SM nanowire is partially covered by the SC shell [Fig.~\ref{Fig:device}(b)], so that the side gate can be used to tune the chemical potential of the SM nanowire.
To date, systems with partially-covered SM nanowires are among the most preferred options in the realization of MZMs in hybrid SM-SC structures~\cite{lutchyn2018majorana}.

As mentioned above, the Majorana-nanowire model requires a large enough Zeeman splitting in the SM nanowire to realize MZMs.
This Zeeman splitting is generated via an applied magnetic field, which might destroy the superconductivity of the SC shell.
Consequently, materials with large Land\'{e} g factors are preferred, to produce a sufficient Zeeman splitting with a comparatively small magnetic field.
In 2020, it was proposed~\cite{vaitiekenas2020flux} that a full-shell SM-SC nanowire, as shown in Fig.~\ref{Fig:device}(c), requires only a tiny parallel magnetic field to reach the topological phase. The magnetic field aims to induce a magnetic flux $\Phi$ threading through the full-shell geometry, rather than a Zeeman splitting, thus the requirement of a large g factor is therefore removed, potentially increasing possible candidates of SM materials. Specifically, the flux $\Phi$ creates fluxoids in the superconducting paring of the SC shell, i.e., $\Delta=|\Delta|e^{in\phi}$, with $\phi$ the angle around the nanowire axis and $n$ the integer closest to the ratio $\Phi/\Phi_0$ with $\Phi_0=h/2e$ being the superconducting flux quantum. It was predicted that MZMs can be engineered in the odd-$n$ cases. Experimentally, for a full-shell nanowire with the SM core diameter of around 130~nm, and the SC shell thickness of 30~nm, the required magnetic field for realizing the $n=1$ topological superconducting regime is about 0.1 T~\cite{vaitiekenas2020flux}.
As another advantage, the full-shell structure naturally protects the SM from impurities and random surface doping, thus enabling a reproducible way of growing SM-SC nanowires with essentially identical electrostatic environments.
As one possible disadvantage, however, full-shell SM-SC nanowires do not allow the control of their electron density via metallic gates, due to the screening of the external electric fields by the SC shell.
Indeed, it is stated that the chemical potential of a full-shell SM-SC nanowire is instead determined by the diameter of the SM core~\cite{vaitiekenas2020flux}.
It is therefore expected to realize MZMs in full-shell SM-SC nanowires with appropriate SM core diameter~\cite{vaitiekenas2020flux}. However, numerical studies~\cite{woods2019electronic} on the electronic structure of full-shell InAs--Al nanowires indicate that this system may be hard to be tuned to the topological phase.
Specifically, it was shown that the spin-orbit coupling, which is crucial in the realization of MZMs, is generically weak for subbands considered in the Majorana-nanowire model, i.e., subbands near the Fermi surface.
Consequently, the topological phase requires a fine-tuned chemical potential, and thus a fine-tuning of the SM core diameter. Though clear ZBCPs robust to the back-gate voltage have been observed within a large scope in the anticipated $n=1$ topological superconducting regime of full-shell hybrid InAs-Al nanowires \cite{vaitiekenas2020flux}, their origin is still under debate. Indeed, a subsequent tunneling spectroscopy experiment~\cite{valentini2021nontopological} on full-shell InAs-Al nanowires demonstrated that ZBCPs arising from topologically trivial Andreev bound states (ABSs) could occur when the tunnel junction is longer than 150 nm to accommodate an interacting quantum dot.     

Based on the Majorana-nanowire model, to drive the SM nanowire into the topological superconducting phase, the Zeeman field should be orthogonal to the Rashba spin-orbit field in the nanowire. In devices discussed above, the Rashba spin-orbit field is mainly determined by the electrostatic potential across the wire section and thus perpendicular to the wire axis \cite{de2018electric,bommer2019spin}. Therefore, the applied magnetic field is required to be parallel to the wire. This requirement is however unsatisfiable for engineering topological nanowire networks, such as topological T-shaped~\cite{alicea2011non} and X-shaped~\cite{zhou2020phase} junctions in which the nanowire segments have different axis directions. As a possible solution, it was proposed~\cite{sau2010generic} to generate a Zeeman splitting in a SM from the exchange interaction between the ferromagnetic insulator (FMI) and the SM. Following this proposal, in 2021, an FMI shell (EuS) was epitaxially grown between the Al shell and the InAs nanowire with a layout shown in Fig.~\ref{Fig:device}(d)~\cite{vaitiekenas2021zero}.
Without an applied magnetic field, in Ref.~\cite{vaitiekenas2021zero} a ZBCP was observed and a remnant effective Zeeman field exceeding 1 T was inferred. Numerical simulations~\cite{liu2021electronic} on the electronic properties of this hybrid InAs/EuS/Al nanowire imply that the magnetic proximity effects at the Al/EuS as well as the InAs/EuS interfaces are both essential to achieve topological superconductivity, without magnetic field applied.

MZMs are also predicted~\cite{hell2017two,pientka2017topological} to exist in a planar Josephson junction (JJ)~\cite{fornieri2019evidence,ren2019topological,dartiailh2021phase} that consists of a 2D SM electron gas covered by two disconnected SC shells [Fig.~\ref{Fig:device}(e)]. The narrow SM stripe uncovered by SC forms the junction region. In the presence of a new tuning knob, i.e., the phase difference of the JJ tuned by the magnetic flux induced by a perpendicular magnetic field $B$, a zero or tiny parallel magnetic field is required to induce Zeeman splitting. Theoretically, the topological superconducting phase can be achieved (with two MZMs localized at the two ends of the SM stripe) even without an applied magnetic field when the phase difference is $\pi$ and the junction is transparent~\cite{pientka2017topological}.
In the presence of normal backscattering in the junction, a weak magnetic field, in addition to the phase difference, is required to induce a Zeeman splitting.
The induced gap in planar JJs is relatively smaller than that of other platforms mentioned above due to its 2D feature: now quasiparticles can travel parallel to the junction, leading to a much longer flight time in the non-superconducting SM region~\cite{laeven2020enhanced}. Fortunately, these parallel quasiparticle trajectories, based on Ref.~\cite{laeven2020enhanced}, can be greatly suppressed in a stripe with zigzag geometry, in which the generated MZMs are expected to be more robust: they are protected by a much larger (one-order larger in amplitude) induced gap and have much smaller localization lengths~\cite{laeven2020enhanced}.

Basically, all the platforms shown in Figs.~\ref{Fig:device}(b)--\ref{Fig:device}(e) can be described by the Majorana-nanowire model~\cite{lutchyn2010majorana,oreg2010helical} in different dimensions with proper modifications. In most of the theoretical works, the relevant parameters, e.g., effective electron mass, Land\'{e} g factor, chemical potential, proximity induced gap, spin-orbit coupling strength, etc., are treated as independent variables in numerical simulations. In real experiments, however, the tunable knobs are the magnetic field and gate voltages, which together determine the values of all other parameters in a complex way. In order to make a connection to realistic experiments, it is suggested to study the hybrid SM-SC systems within the framework of Schr\"{o}dinger-Poisson method~\cite{mikkelsen2018hybridization,antipov2018effects,woods2018effective}, under the Thomas-Fermi approximation~\cite{wojcik2018tuning,winkler2019unified,escribano2019effects,escribano2020improved,liu2021electronic}.
This is a systematic and close-to-experiment numerical approach that takes into account all experimentally relevant effects, e.g., the orbital effect induced by a magnetic field, superconducting proximity effect, SM-SC work function difference, and electrostatic environment.

\section{Recent progress on detecting MZMs}\label{Sec III}
In addition to engineering MZMs, detecting MZMs in SM-SC heterostructures is another challenging task. In this section, we review the latest progress on the detection of MZM signatures. These detection methods include (i) the tunneling spectroscopy detection of the existence of MZMs in Sec.~\ref{sec:tunneling_spectroscopy}; (ii) the detection of MZM non-locality in Sec.~\ref{extension}; (iii) the detection of MZM signature from ac Josephson effect in Sec.~\ref{sec:4pi_JJ}; (iv) the detection of MZMs electron-hole weights with Coulomb oscillation in Sec.~\ref{sec:cb_oscillation}, and (v) nonlocal tunneling spectroscopy measurement in Sec.~\ref{sec:three-terminal}. 
In addition to the experimental efforts above, other potential detection protocols (e.g., topological Kondo effect~\cite{BeriCooperPRL12,AltlandBeriPRL14,AffleckJournalStatMech13,MartinCooperPRB14,AltlandIOP14,ErikssonPRB14,ErikssonPRL14,ZazunovIOP14,liu2021topological}) can be found in Ref.~\cite{zhang2019next}.

\subsection{Tunneling spectroscopy}
\label{sec:tunneling_spectroscopy}
As one of the earliest introduced techniques, tunneling spectroscopy remains among the most popular tools to detect MZMs in SM-SC heterostructures and other topological SC candidates. Theoretically, MZMs are exotic zero-energy excitations with equal electron and hole components. Because of this feature, resonant tunneling into a MZM is predicted to arrive at a quantized ZBCP with a height of $2e^2/h$ at zero temperature~\cite{sengupta2001midgap,law2009majorana,flensberg2010tunneling,wimmer2011quantum}. This quantized ZBCP, being protected by topology, is predicted as robust against perturbations, e.g., the tuning of magnetic field and gate voltages, as long as the topology remains invariant.

However, since the first report of ZBCP in 2012, most of the observed ZBCPs~\cite{deng2012anomalous,das2012zero,finck2013anomalous,churchill2013superconductor,deng2016majorana,chen2017experimental,suominen2017zero,gul2018ballistic,sestoft2018engineering,vaitiekenas2018effective,deng2018nonlocality,de2018electric,bommer2019spin,grivnin2019concomitant} are much smaller than the predicted quantized value, except for some recently reported nearly-quantized ZBCPs~\cite{nichele2017scaling,pan2022situ,zhang2021large,heedt2021shadow,song2021large,wang2022observation}.
Meanwhile, some other predicted MZM signatures remain missing in most tunneling spectroscopy measurements. These missing features include (i) the predicted correlation between the emergence of ZBCPs and the close-and-reopen of the superconducting gap gap~\cite{lutchyn2010majorana,oreg2010helical}; and (ii) the predicted oscillations of MZM splitting energy as a function of magnetic field or chemical potential, dubbed as Majorana oscillations~\cite{sarma2012splitting,rainis2013towards}.
For issue (i), it is explained~\cite{stanescu2012to} that the gap close-and-reopen is a bulk property of topological phase transition that is not necessarily visible in the tunneling spectroscopy. Specifically, in the topologically trivial phase, the lowest-energy states are delocalized (localized) through the wire when the chemical potential is lower (larger) than a critical value that is of the order of the proximity-induced superconducting gap, resulting in an invisible (visible) gap closure in the tunneling spectroscopy which is related to the local density of states at the end of the wire. We mention that a recent experiment reported the concomitant opening of a bulk gap with emerging ZBCPs.
The Majorana oscillation [issue (ii)], on the other hand, is explained as being sabotaged by either the inverse proximity effect arising from drain-MZM coupling~\cite{danon2017conductance},
or the Coulomb interaction between MZMs and the image charges in the dielectric material around the SM nanowire~\cite{dominguez2017zero}.

Despite the explanations on the imperfect theory-experiment agreements, the frequent observation of non-quantized ZBCPs in nanowire experiments since 2012 has stimulated many theoretical works on exploring the possible origins of ZBCPs, see, e.g., Refs.~\cite{pientka2012enhanced,prada2012transport,liu2012zero,kells2012near,pikulin2012zero,rainis2013towards,roy2013topologically,stanescu2013disentangling,liu2017andreev,reeg2018zero,moore2018quantized,vuik2019reproducing,pan2021quantized,cayao2021confinement}. Most recently, tunneling spectroscopy experiments of nanowires~\cite{chen2019ubiquitous,junger2020magnetic} have revealed non-Majorana ZBCPs owing to different mechanisms. At present, these progress pose a serious doubt that all observed ZBCPs in SM-SC heterostructures might originate from topologically trivial zero-energy ABSs, see Ref.~\cite{prada2020andreev} for a review. Indeed, following thorough numerical simulations, Ref.~\cite{pan2020physical} further demonstrates that most of the experimental ZBCPs in InAs and InSb nanowire devices might have been induced by strong disorders in the chemical potential, proximity-induced superconducting gap, or effective Land\'{e} g factor.
It is worth mentioning that smooth-potential-induced topologically trivial quasi-MZMs~\cite{vuik2019reproducing,moore2018two,pan2020physical} can also lead to quantized ZBCP plateaus~\cite{moore2018quantized}.
Both MZMs and quasi-MZMs stand out from ABS interruptions, as the former ZBCPs remain under the tuning of multiple experimental parameters.
Notably, the observation of ZBCP plateaus in real experiments is difficult as it depends on experimental details, e.g., temperature, tunnel barrier, and contact resistance, although the unknown contact resistance is expected as avoidable in a four-terminal device~\cite{song2021large}. Nevertheless, a recent experiment~\cite{wang2022observation} observed plateau regions for ZBCPs within $5\%$ of the quantized conductance value $2e^2/h$ when individually sweeping back-gate voltage, tunnel-gate voltage, and parallel magnetic field. This progress represents a further advance towards the detection of MZMs or quasi-MZMs.

\begin{figure}[t!]
\centering
\includegraphics[width=\columnwidth]{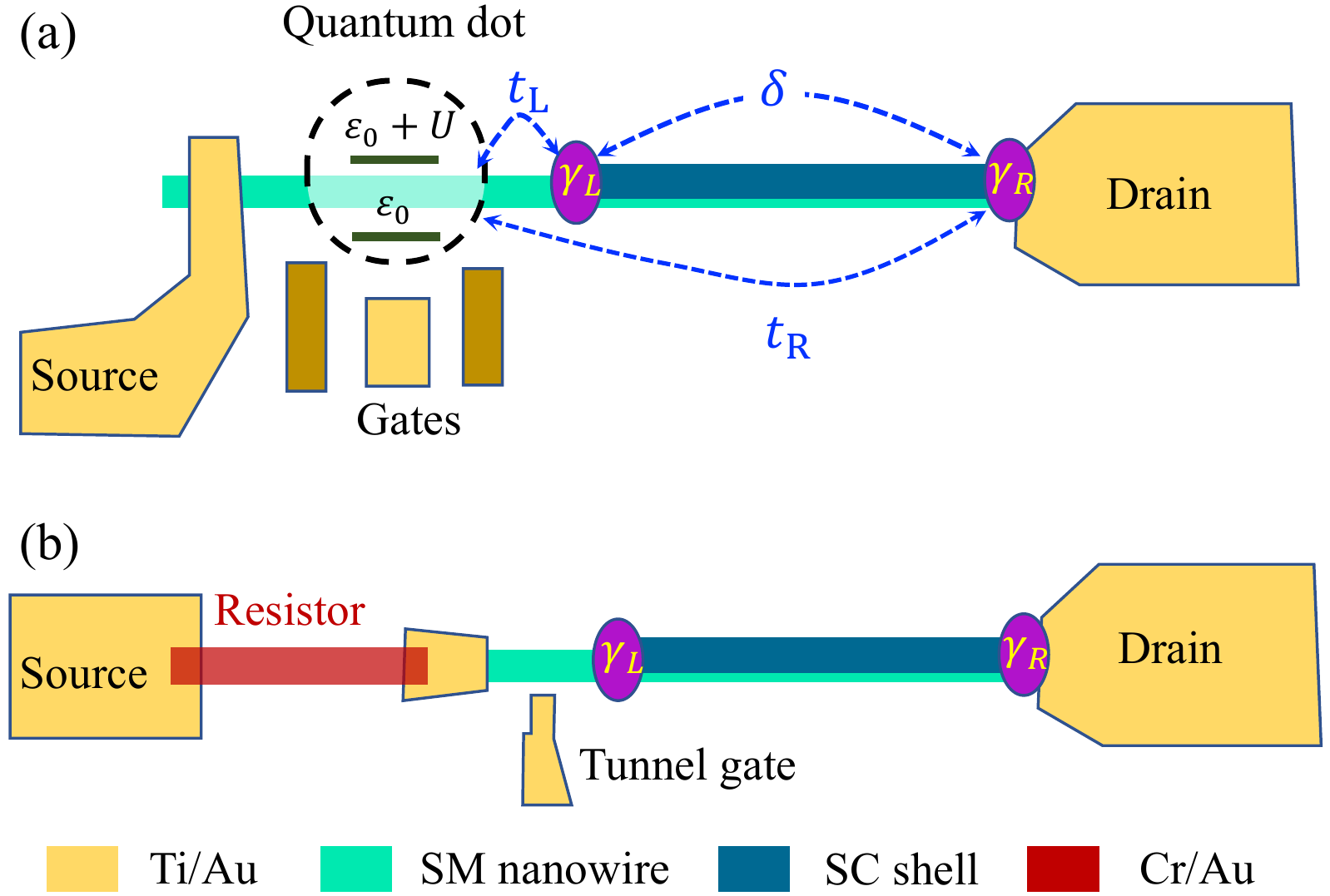}
\caption{Schematics of the extended tunneling spectroscopy. (a) An intentional quantum dot formed between the source lead and the hybrid SM-SC nanowire enables the measurement of the nonlocality of MZMs. The quantum dot, with an energy level $\varepsilon_0$ and a charging energy $U$, couples to the left and right MZMs with tunneling amplitudes $t_L$ and $t_R$, respectively, and the hybridization energy between the two MZMs is $\delta$. (b) Tunneling spectroscopy measurement in the presence of a resistor in series with the source lead.
The resistor contains bosonic modes that dissipate energy during transport processes.
Transport is thus suppressed by coupling to the resistor:
this phenomenon is known as dynamical Coulomb blockade.
}\label{Fig:dot}
\end{figure}

\subsection{Extensions of tunneling spectroscopy}\label{extension}
It was realized that tunneling spectroscopy protocol with minor modifications can provide alternative insights (other than ZBCPs) to detect MZMs.
For instance, a quantum dot that locates at the end of the nanowire as shown in Fig.~\ref{Fig:dot}(a) is predicted to detect the nonlocality of MZMs~\cite{prada2017measuring,clarke2017experimentally}, while an Ohmic resistance in series with the electrode~\cite{liu2013proposed,liu2022universal,zhang2022supressing} as shown in Fig.~\ref{Fig:dot}(b) could confirm the existence of MZMs via its unique universality class.

To start with, nonlocality is a unique property that can distinguish MZMs from ABSs.
Intuitively, tunneling spectroscopy obtained from a local probe is unable to reflect the nonlocal property. Surprisingly, however, the nonlocality is proposed to be measurable with the assistance of a quantum dot ~\cite{prada2017measuring,clarke2017experimentally}, in a device schematically shown in Fig.~\ref{Fig:dot}(a). As indicated, there exists five energy scales, i.e., the dot level $\varepsilon_0$, the Coulomb charging energy $U$, the dot-MZM couplings $t_L$ and $t_R$, and the inter-MZM coupling $\delta$. In experiments, $U$ is inversely proportional to the dot size and $\varepsilon_0$ can be freely tuned by the voltage gate near the dot. Note that $U=0$ also works for measuring nonlocality as analyzed in Ref.~\cite{clarke2017experimentally}. The hierarchy of the rest three energy scales characterizes a pair of well-separated MZMSs, an ABS, or the cases in between. For a pair of MZMs each of which is localized at the opposite ends of a finite-length nanowire [see Fig.~\ref{Fig:dot}(a)], both $\delta$ and $t_R$ decay exponentially as a function of the wire length. In contrast, for an ABS localized at the left wire end, so are its two Majorana components, $t_L$ equals $t_R$ and $\varepsilon_0$ can be an arbitrary value including zero. The ratio of $\sqrt{t_R/t_L}$ has been numerically demonstrated to be an estimator that can quantify the degree of nonlocality of a subgap bound state~\cite{prada2017measuring}. Theoretically, different amplitudes of $\delta$, $t_L$, and $t_R$ lead to different characteristic tunneling spectroscopy patterns, from which the quantity $\sqrt{t_R/t_L}$ can be extracted, as predicted in Ref.~\cite{prada2017measuring}. This protocol has been implemented in an experiment~\cite{deng2018nonlocality} based on hybrid InAs-Al nanowires, in which either a moderate degree of nonlocality, suggesting a partially separated ABS, or a highly nonlocal subgap state consistent with a pair of well-separated MZMs, were observed.

Previously, to observe quantized ZBCPs, noble metals (e.g., Au and Pt) with negligible resistances are commonly chosen as lead materials (see e.g., Refs.~\cite{deng2016majorana,nichele2017scaling, gul2018ballistic,vaitiekenas2018effective,zhang2021large,song2021large,wang2022observation}). 
The authentication of the quantized ZBCP signature, as has been discussed above, is however undermined by possible fake positive signals arising from disorder-induced ABSs.
Fortunately, these fake positive signals can be suppressed by attaching a resistor (dissipation) to the leads~\cite{liu2013proposed,liu2022universal}, as schematically shown in Fig.~\ref{Fig:dot}(b).
The real MZM signal remains, however, for a resistor with a proper impedance.

Briefly, a resistor can be considered as a bath that contains bosonic modes. During tunneling events, these modes can be excited to higher-energy levels, during which energies are dissipated~\cite{grabert2013single}.
This phenomenon, known as dynamical Coulomb blockade, suppresses tunnelings at low enough energies.
Most importantly, dissipation influences the tunneling into a MZM and an ABS in different ways~\cite{liu2013proposed,liu2022universal}, leading to distinct universal or non-universal features.
One can thus distinguish the MZM signal from the ABS ones by analyzing the transport features at low enough energies, counter-intuitively, with dissipation intentionally included in the lead.
As the core idea, dissipation effects introduce interaction between transported electrons and environmental bosons~\cite{grabert2013single} and cause renormalization effects of electron tunnelings as running the system energy scale. Importantly, due to this renormalization effect, Majorana and topologically trivial transports show completely different scaling behaviors (in both temperature and voltage bias) that suffice to tell MZM from trivial bound states.
The idea to suppress ABS-induced ZBCPs has been recently verified experimentally~\cite{zhang2022supressing} when the dissipation is strong enough ($R \ge 0.21 h/e^2$).
With a smaller dissipation in the sample ($R \approx 0.1 h/e^2$), the ABS-induced ZBCP is not fully suppressed at experimentally accessible temperatures ($T > 24$ mK) of Ref.~\cite{wang2022large}. Instead, the ABS-induced ZBCP now displays a clear and interesting competition between dissipation and thermal fluctuations.

\begin{figure}[t!]
\centering
\includegraphics[width=\columnwidth]{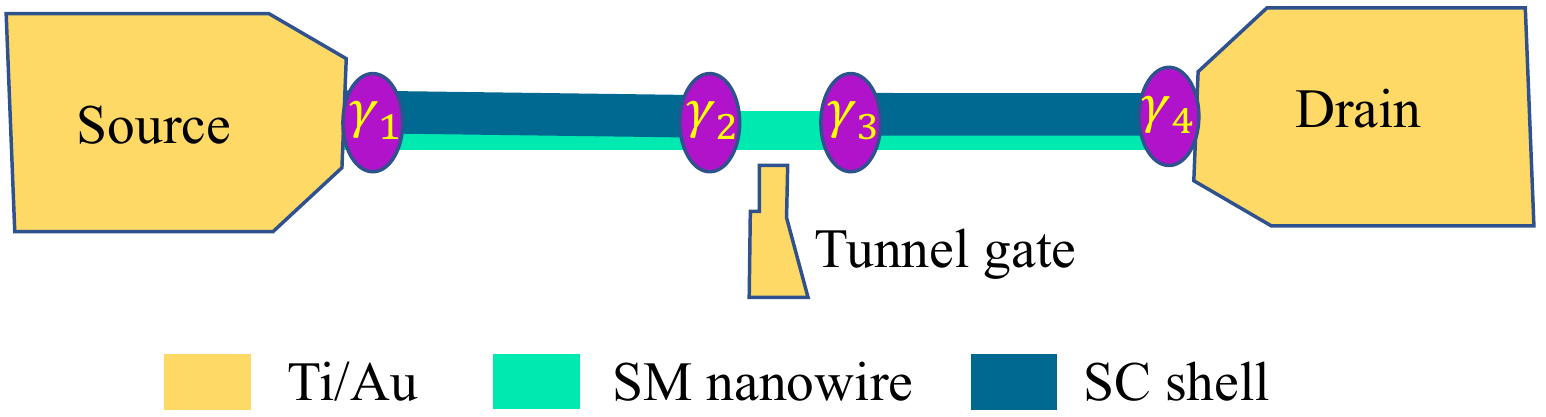}
\caption{(Color online) Schematic JJ based on a hybrid SM-SC nanowire. The tunnel gate tunes the junction transparency. If each nanowire segment on the two sides of the junction hosts two well-separated MZMs, the fractional ac Josephson effect is predicted to be measurable in such a device with a source-drain bias voltage and low junction transparency.}\label{Fig:JJ}
\end{figure}

\subsection{Fractional Josephson effect}
\label{sec:4pi_JJ}
Josephson effect~\cite{josephson1962possible,josephson1965supercurrents} is a unique transport phenomenon across a JJ.
Experimental detection of MZMs also includes the probing of fractional Josephson effect~\cite{kwon2004fractional} in SM-SC nanowire setups~\cite{rokhinson2012fractional,tiira2017magnetically,kamata2018anomalous,laroche2019observation} or HgTe-based topological insulator setups~\cite{wiedenmann20164pi,bocquillon2017gapless,deacon2017josephson}. A hybrid SM-SC nanowire JJ is schematically shown in Fig.~\ref{Fig:JJ}, where a fraction of the SC shell in the middle is etched to create a JJ. A nearby tunnel gate is used to tune the junction transparency.

When the hybrid SM-SC nanowire in Fig.~\ref{Fig:JJ} is driven to the topological superconducting phase, the left (right) hybrid nanowire segment supports two MZMs $\gamma_{1,2}$ ($\gamma_{3,4}$) at the ends.
With a long-enough nanowire, MZMs sufficiently decouple, and
the two outer MZMs ($\gamma_{1,4}$) are projected out of the low-energy subspace of the topological JJ.
In this case, the junction Hamiltonian effectively becomes
$H_\textrm{eff}(\phi)=t(\phi)(\hat{n}-1/2)$ where
$t(\phi)$ is the inter-MZM coupling as a function of the phase difference $\phi$ across the junction, and $\hat{n}=(1+i\gamma_2\gamma_3)/2$ is the number operator. Its eigenvalues are conserved since $[\hat{n},H_\textrm{eff}]=0$.
The eigenvalues $n=0$ and $n=1$ correspond to the state of even and odd parities, respectively.
Following a simple symmetry analysis~\cite{marra2022bMajorana}, $t(\phi)$ is $4\pi$-periodic in $\phi$.
Consequently, at zero temperature, the equilibrium %the zero-temperature
dc Josephson current mediated by a pair of MZMs becomes $I(\phi)=-\frac{2e}{\hbar}(n-1/2) \partial_\phi t( \phi)$, which is also $4\pi$-periodic in $\phi$~\cite{kitaev2001unpaired}.
This $4\pi$ periodicity is considered as a MZM signature, as it is in strong contrast to the $2\pi$ periodicity in conventional JJs.
Physically, the doubling of periodicity reflects the fractionalization of a MZM: it can be considered as `half' of a regular fermion.

In real experiments, this $4\pi$-periodic feature however encounters multiple challenges~\cite{fu2009josephson,lutchyn2010majorana,sanjose2012ac}.
To begin with, as the prerequisite of the $4\pi$-periodic feature, parity conservation unfortunately can be violated by e.g., the hybridization between the inner and outer MZMs
or the inelastic particle transition that involves
the quasiparticle continuum above the gap (i.e., the quasiparticle poisoning).
Fortunately,
these detrimental effects
are less severe in
ac Josephson junctions~\cite{sanjose2012ac,houzet2013dynamics,dominguez2012dynamical,sau2017detecting}.
Detection of the fractional ac Josephson effect has been performed by measuring either the Shapiro steps~\cite{shapiro1963josephson} or the Josephson frequency $f$.
Specifically, when influenced by a microwave with frequency $\Omega$, the dc component of the current across a low-transparency topologically trivial JJ contains
the Shapiro steps at biases $V_n=nV_0$, for all positive $n$ and $V_0=h\Omega/2e$~\cite{shapiro1963josephson}.
To date, Shapiro steps at $V=V_0,~2V_0,~3V_0,~4V_0$ have been observed for JJ in the trivial regime~\cite{rokhinson2012fractional}.
In the topological regime, by contrast, only even numbers of $n$ are allowed theoretically~\cite{houzet2013dynamics,dominguez2012dynamical,sau2017detecting}. This is, however, not observed in a real experiment~\cite{rokhinson2012fractional}, where the Shapiro step at $V=V_0$ is absent while the one at $V=3V_0$ still remains.
This theory-experiment discrepancy has been visited by multiple works~\cite{pikulin2012phenomenology,dominguez2012dynamical,virtanen2013microwave,de2016interplay,sau2017detecting,pico2017signatures,le2019joule} after taking more experimental details into consideration.
As another option, the fractional Josephson effect can be detected by measuring the Josephson frequency. Indeed, in the topological regime, the Josephson frequency is predicted~\cite{sanjose2012ac} to be half of that in the trivial regime.
This phenomenon has been observed by Ref.~\cite{kamata2018anomalous}. However, all Shapiro steps $V=nV_0$ are observed in Ref.~\cite{kamata2018anomalous}, in contrast to the topological criteria introduced above.
The halved Josephson frequency was interpreted as trivial changes in the superconducting circuit surrounding the device strongly affected by the applied magnetic field~\cite{kamata2018anomalous}.
The halved Josephson frequency has also been reported in a JJ made of hybrid InAs-Al nanowire~\cite{laroche2019observation}.

\begin{figure}[t!]
\includegraphics[width=\columnwidth]{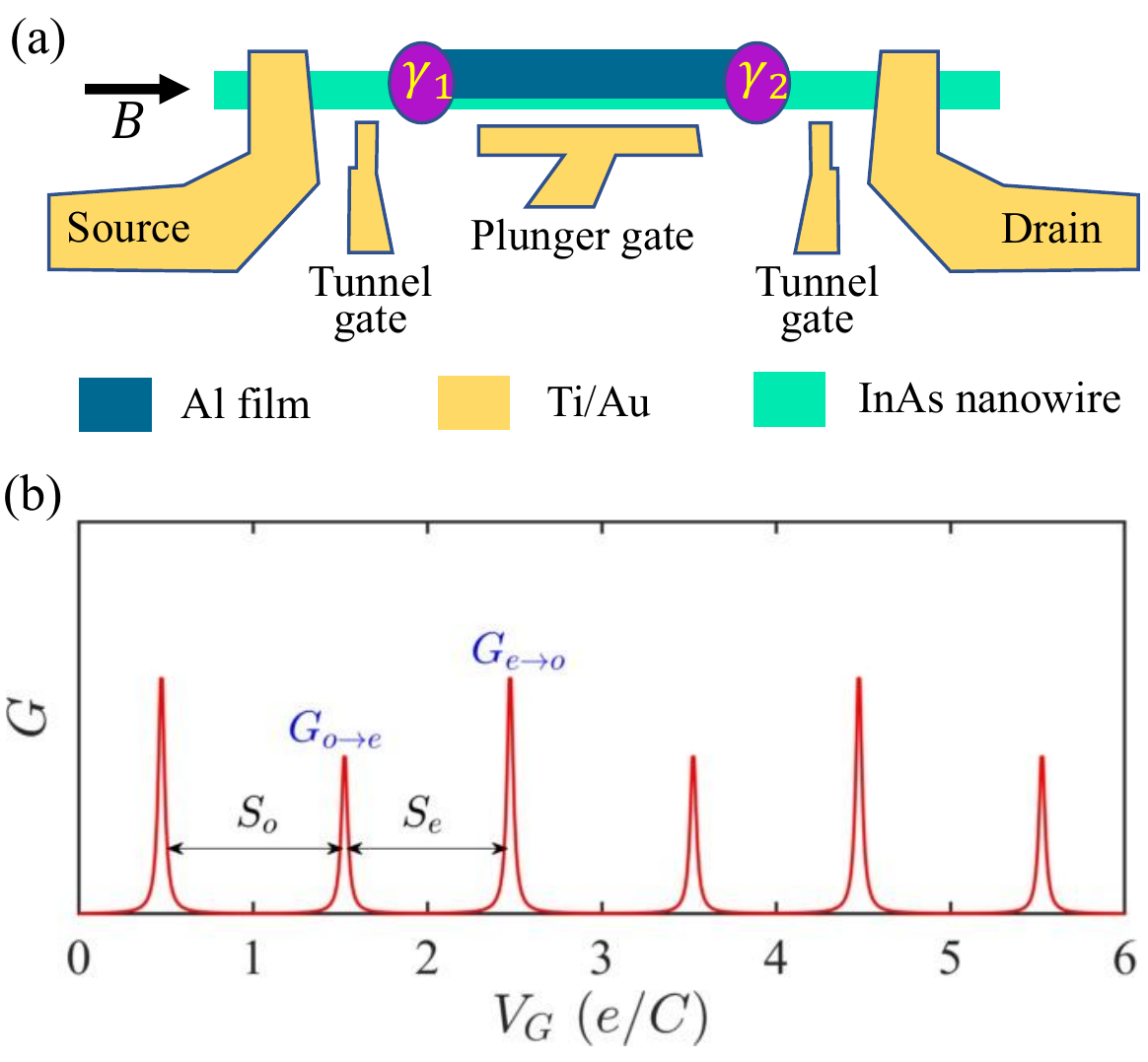}
\caption{(Color online) (a) A schematic device for measuring Coulomb blockade transport through a floating SM-SC nanowire island. Applying a parallel magnetic field $B$ may induce a pair of MZMs at the two ends of the island. The bias voltage $V_{SD}$ drives a charge current $I$ through the island, gate voltages $V_{T1}$, $V_{T2}$ control the tunnel barriers between the metallic leads and their nearest island ends, and $V_G$ controls the island electron number through adjusting the electrostatic energy of the nanowire island. (b) Typical Coulomb-blockade conductance peaks. Peak spacings for odd (o) and even (e) Coulomb valleys are indicated as $S_o$ and $S_e$, respectively. Peak heights at odd-to-even (even-to-odd) parity transitions occurring upon increasing $V_G$ are denoted as $G_{o\rightarrow e}$ ($G_{e\rightarrow o}$). In experiments, both the peak spacings and peak heights oscillate as $B$ various.}\label{Fig:island}
\end{figure}

\subsection{Coulomb blockade conductance oscillation}
\label{sec:cb_oscillation}
Coulomb blockade refers to the phenomenon where charge transport is suppressed by the Coulomb interaction. As known, Coulomb blockade transport can provide quantitative information about the charge and energy spectrum of a mesoscopic island~\cite{von2001spectroscopy,aleiner2002quantum}. It thus also helps the detection of MZMs, especially their non-local features.

Figure \ref{Fig:island}(a) shows a common structure with Coulomb blockade involved. In experiments~\cite{higginbotham2015parity,albrecht2016exponential,sherman2017normal,albrecht2017transport,vaitiekenas2018selective,o2018hybridization,shen2018parity,shen2021full,bjergfelt2021superconductivity,estrada2022excitations,valentini2022majorana} with devices of the considered structure, a source-drain bias voltage is applied to drive a charge current through a floating hybrid SM-SC nanowire. This finite-size island has a charging energy $E_C$ and a proximity-induced superconducting gap $\Delta$ (at $B=0$).
The plunger gate voltage $V_G$, applied beneath the island, controls the number of electrons on the island and hence the number parity of the island.
The couplings between the island and two metallic leads are controlled by two tunnel gates.
In addition to the tunneling amplitudes tuned by the tunnel gates, the island-lead communications also depend on the relative amplitudes of other relative energy scales including $E_C$, $\Delta$, and the ABS (or MZM) energy.

Assuming that the superconducting island hosts an ABS with energy $E_0$ and finite wavefunction distributions on both ends of the island. The most interesting situation occurs when $E_0<E_C<\Delta$, where the parity of the ground-state of the island can alternate between even and odd as the voltage $V_G$ changes~\cite{van2016conductance}.
At a vanishingly small bias voltage and low temperatures, generally, the transport of electrons through the island is forbidden [the conductance valley shown in
Fig.~\ref{Fig:island}(b)] due to blockade from Coulomb-induced charging energy (i.e., the Coulomb blockade).
By contrast, with a fine-tuned $V_G$, island states with different parities share the same energy.
This energy degeneracy of the island enables resonant transport through the island, leading to the sharp ZBCP shown in Fig.~\ref{Fig:island}(b).
The conductance peak heights associated with the even-to-odd ($G_{e\rightarrow o}$) and odd-to-even ($G_{o\rightarrow e}$) ground-state parity transitions are different, due to the fact that they are related to the electron-like and hole-like components of the ABS (or MZM) of the island. Consequently, the electron-hole weights of the ABS (or MZM) can be extracted from the positions and heights of the consecutive Coulomb blockade conductance peaks~\cite{hansen2018probing}.

By a rate-equation analysis, in the case where the tunnel couplings at both ends of the nanowire are equal, it was found that~\cite{hansen2018probing}
\begin{eqnarray}
\Lambda=\frac{G_{e\rightarrow o}}{G_{e\rightarrow o}+G_{o\rightarrow e}}=\frac{\frac{|u_L|^2|u_R|^2}{|u_L|^2+|u_R|^2}}{\frac{|u_L|^2|u_R|^2}{|u_L|^2+|u_R|^2}+\frac{|v_L|^2|v_R|^2}{|v_L|^2+|v_R|^2}},
\label{eq:lambda}
\end{eqnarray}
where $|u_{L/R}|^2=\sum_\sigma|u_{L/R,\sigma}|^2$, $|v_{L/R}|^2=\sum_\sigma|v_{L/R,\sigma}|^2$, with $u_{L/R,\sigma}$, $v_{L/R,\sigma}$ the spin-resolved local Bogoliubov-de Gennes amplitudes of the ABS at the left/right end of the island. If the island has a space-inversion symmetry, i.e., $|u_L|=|u_R|=|u|$ and $|v_L|=|v_R|=|v|$, Eq.~\eqref{eq:lambda} simplifies to
\begin{equation}
\Lambda=\frac{G_{e\rightarrow o}}{G_{e\rightarrow o}+G_{o\rightarrow e}}=\frac{|u|^2}{|u|^2+|v|^2}. \label{ratio}
\end{equation}
Clearly, in this case, the electron and hole weights can be obtained by measuring the $\Lambda$ of Eq.~\eqref{ratio} in experiments.

The ABS (or MZM) energy $E_0$, on the other hand, can be extracted from
\begin{equation}
\pm E_0=\eta \langle S_{e(o)}\rangle-E_C, \label{E0}
\end{equation}
that relates $E_0$ to the charging energy $E_C$ and the peak spacing average $\langle S_{e(o)}\rangle$ [see Fig.~\ref{Fig:island}(b)].
The charging energy and the lever arm $\eta$ can be estimated from the Coulomb blockade diamonds features.

Equations~\eqref{ratio} and \eqref{E0} address how the information of an ABS (or a MZM) in an SM-SC nanowire island can be obtained from Coulomb blockade transport features.
The obtained values of $\Lambda$ and $E_0$, which are magnetic-field dependent in real experiments~\cite{o2018hybridization,shen2018parity,shen2021full}, can help to distinguish a MZM from ABSs, as the Coulomb blockade transport mediated by a pair of MZMs is predicted to be distinct from that by an ABS. More specifically, the MZMs mediated Coulomb blockade transport has the following unique features: (i) $E_0$ decreases exponentially with increasing the wire length \cite{sarma2012splitting}. (ii) $E_0$ and $\Lambda$ are pined to 0 and $1/2$, respectively, for long wires in which $|u|=|v|$ is guaranteed. For short wires with a finite MZM overlap, both $E_0$ and $\Lambda$ oscillate with the magnetic field $B$. The oscillating amplitude and period are both larger under a greater magnetic field, as long as the system remains in the topological regime~\cite{hansen2018probing}.
(iii) The oscillations of $E_0$ and $\Lambda$ have a $\pi/2$ phase shift~\cite{hansen2018probing}.
Features akin to theoretical predictions of feature (i) have been claimed to be observed experimentally~\cite{albrecht2016exponential} by measuring several devices with different wire lengths.
In Ref.~\cite{albrecht2016exponential}, however, the observed oscillation of $E_0$ decays when $B$ increases, in strong conflict with the predicted feature (ii).
In some other experimental efforts~\cite{o2018hybridization,shen2021full}, the observed oscillation of $E_0$ also decays when $B$ increases, but the $\pi/2$ phase shift mentioned in feature (iii) has been observed. Both features (ii) and (iii) are however absent in Ref.~\cite{shen2018parity}. Several theoretical works \cite{chiu2017conductance,dmytruk2018suppression,fleckenstein2018decaying,cao2019decays} tried to explain these discrepancies with different assumptions. However, the discrepancies can not be fully explained by these theories, except for Ref.~\cite{cao2019decays}, which assumes that there might exist steplike Rashba spin-orbit couplings along these nanowire islands, due to the potentially nonuniform electrostatic potential at the center and the two sides of the nanowires.

Before closure, we also mention that with the dissipation effect present as shown in Sec.~\ref{extension}, the tunneling signatures through a Coulomb blockade island will be modified~\cite{liu2020revealing}; and this proposal provides a hallmark for the nonlocal coherent nature in the Fu-teleportation~\cite{fu2010electron,semenoff2006teleportation,hutzen2012Majorana}, which are different in comparison to the incoherent transports from other non-topological Coulomb-blockade systems.

\begin{figure}[t!]
\includegraphics[width=\columnwidth]{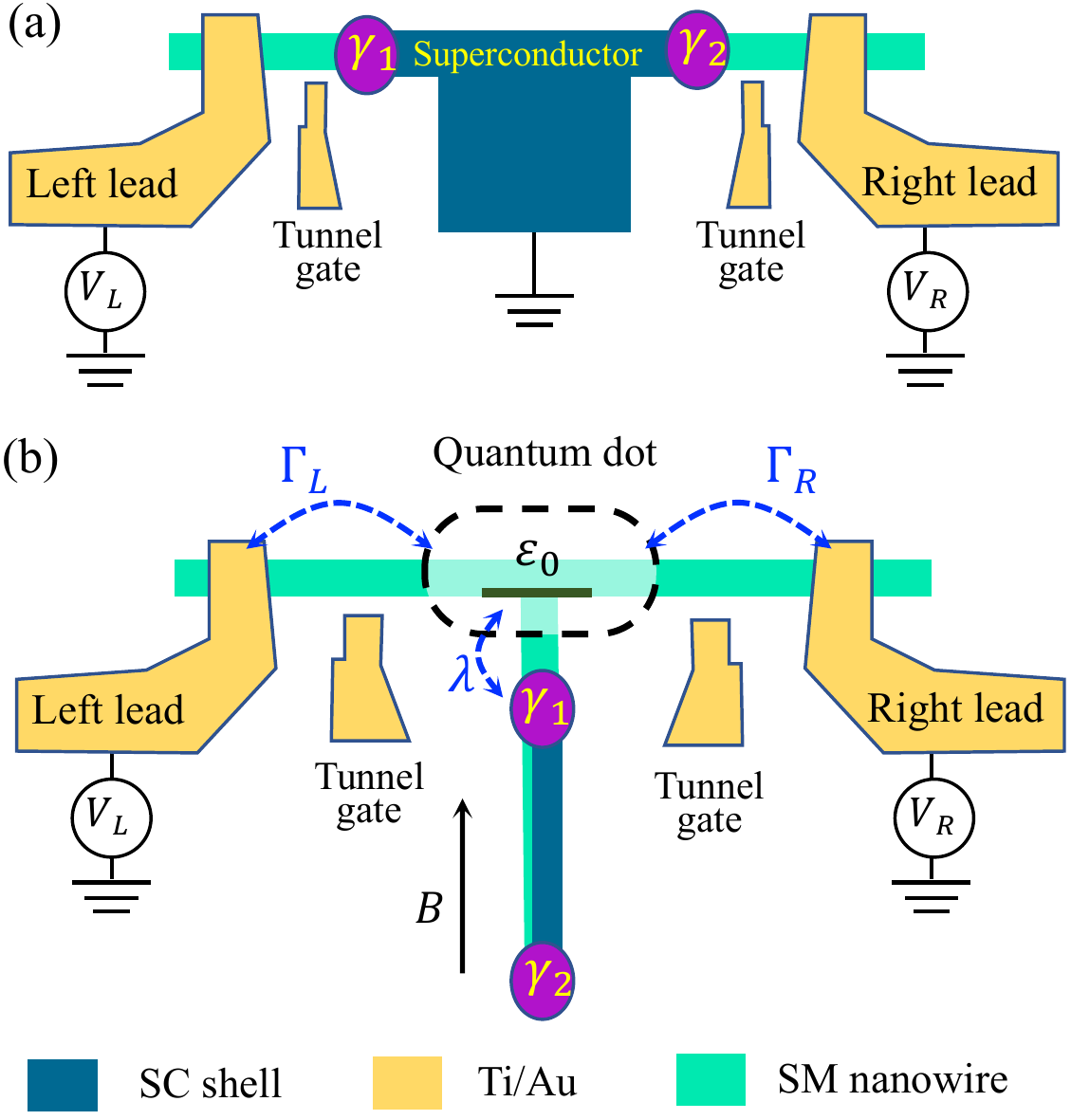}
\caption{(Color online) (a) Schematic of a three-terminal transport device. Two normal leads are connected to a central grounded superconducting region, where the tunnel barriers at two ends can be controlled by tunnel gates. (b) Schematics of a system comprising a spinless quantum dot coupled to one end of a floating hybrid SM-SC nanowire. The conductance through the dot is measured by adding two external leads. }\label{Fig:3terminal}
\end{figure}

\subsection{Three-terminal transport}
\label{sec:three-terminal}
As a more direct option~\cite{rosdahl2018andreev,pan2021three,hess2021local,pikulin2021protocol}, the nonlocal nature of MZMs can be revealed by measuring both local and nonlocal tunneling spectroscopy in a three-terminal device, as schematically shown in Fig.~\ref{Fig:3terminal}(a).
Briefly, as MZMs are predicted as correlated states localized at two ends of a hybrid SM-SC nanowire, a strong correlation is anticipated between local tunneling spectroscopy measured at each ends~\cite{sarma2012splitting}, even when two MZMs fully decouple.
This criterion is preferred to be carried out in experiments with long enough hybrid SM-SC wires. Indeed, correlation might as well appear in topologically trivial systems, if the wire is shorter than the localization length of regular fermionic states or ABSs~\cite{hess2021local}.

The nonlocal conductance spectroscopy employs two spatially separated leads to probe three global properties: the bulk superconducting gap, the induced gap, and the induced coherence length~\cite{rosdahl2018andreev}.
Among them, the bulk superconducting gap is barely a benchmark of the presence of superconductivity.
With the presence of the bulk superconductivity, the second measured quantity, i.e., the induced gap conveys the information of possible topological phase transition:
near the transition point, the induced gap is predicted to experience a gap closure and reopen.
The observed topological phase transition, if accompanied by a strong correlation of local conductance at both ends of the nanowire, is currently believed as a strong evidence of the presence of MZMs.
Following this protocol, some three-terminal experiments~\cite{anselmetti2019end,puglia2021closing,menard2020conductance,wang2022parametric,poschl2022nonlocal,banerjee2022local} have recently been carried out in SM-SC heterostructures.

As the latest experimental progress~\cite{aghaee2022inas}, the Microsoft group realizes the non-local tunnel measurement, which is believed as a more trustworthy protocol than the standard tunneling spectroscopy method (shown in Sec.~\ref{sec:tunneling_spectroscopy}) for the MZM detection. More specifically, Ref.~\cite{aghaee2022inas} claims to have detected the topological superconducting phase transition following the three-terminal transport protocol.
This protocol starts with a fast-detection~\cite{pikulin2021protocol} of local tunneling signal at both sides of the wire.
By doing so, they narrow down the phase space to a selected area with ZBCP observed at both ends of the wire. Finally, careful non-local measurement is taken within the selected area, through which a topological phase transition (in the bulk) and the corresponding topological regime have been identified.
Remarkably, three (out of five) devices have passed this protocol~\cite{aghaee2022inas}.
Although its data quality (e.g., the ZBCP height) requires further improvement, Ref.~\cite{aghaee2022inas} is highly inspiring and encouraging to the Majorana hunting community.

In addition to the protocol above, a recent theory~\cite{danon2020nonlocal} of the three-terminal setup predicts that the ratio between the symmetric and asymmetric components of the nonlocal conductance is directly related to the local BCS charges (defined as $Q_{L/R}=|u_{L/R}|^2-|v_{L/R}|^2$) of the bound state close to the left and right normal leads. Essentially, the measurement of local BCS charges $Q_{L/R}$ (or equivalently, electron-hole weights $|u_{L/R}|^2$ and $|v_{L/R}|^2$) can help detecting MZMs~\cite{hansen2018probing,cao2022probing}: For a pair of well-separated MZMs localized at the two ends of a hybrid nanowire, there exists $|u_{L/R}|^2=|v_{L/R}|^2$ and $Q_{L/R}=0$. This protocol of detecting local BCS charge in a three-terminal setup has been tested in a recent experiment~\cite{menard2020conductance}, but no stable zero BCS charge has been observed.

Given the ability of fabricating multi-terminal hybrid SM-SC nanowires and quantum dots therein, another three-terminal protocol, as shown in Fig.~\ref{Fig:3terminal}(b), has been proposed to detect MZM.
As another related proposal, Ref.~\cite{liu2011detecting} studies the system where a resonant level couples to one localized MZM of a floating hybrid SM-SC nanowire.
In this case, the presence of MZM blocks half of the tunneling channels through the resonant level, leading to a ZBCP with height $e^2/2h$ at zero temperature.
As a possible extension of Ref.~\cite{liu2011detecting}, it is proposed to stabilize the MZM signal by introducing a dissipative resonant level system that couples to the partner of the MZM under detection~\cite{ZhangBarangerPRB20,ZhangSpanslattPRB20}.
Briefly, the coupling between the MZM pairs, which is predicted to sabotage the MZM signal~\cite{liu2011detecting}, will be suppressed by the presence of dissipation, thus manifesting the $e^2/2h$ MZM-featured conductance.

\section{Material aspects}\label{Sec IV}
\subsection{Material choice}
The mostly studied epitaxial hybrid SM-SC nanowires thus far are InAs-Al or InSb-Al~\cite{lutchyn2018majorana}. A few works~\cite{sestoft2018engineering,mayer2020superconducting,moehle2021insbas} also studied a similar combination, InAsSb-Al, since it was predicted~\cite{winkler2016topological} that InAsSb has a much stronger intrinsic spin-orbit coupling than InAs and InSb. To enhance the induced SM superconducting gap and its tolerance to magnetic field and thermal fluctuation, $s$-wave SCs Sn and Pb have been tested in the fabrication of hybrid InSb-Sn~\cite{pendharkar2021parity}, InSb-Pb~\cite{jung2021universal}, and InAs-Pb~\cite{kanne2021epitaxial} nanowires. Tunneling spectroscopy measurements show that such SM-SC nanowires based on Sn or Pb have larger induced superconducting gaps, critical magnetic fields, and higher critical temperatures than InAs-Al and InSb-Al nanowires, see Table \ref{scpara}.

Despite the above advantages from superior SCs, InAs and InSb nanowires were predicted to commonly suffer from disorder interruptions that can potentially sabotage the detection of MZMs (see, e.g., Refs.~\cite{pan2020physical,sarma2021disorder}).
Indeed, these disorders, either long or short-ranged, are known to induce topologically trivial ZBCPs~\cite{pientka2012enhanced,prada2012transport,liu2012zero,kells2012near,pikulin2012zero,rainis2013towards,roy2013topologically,stanescu2013disentangling,liu2017andreev,reeg2018zero,moore2018quantized,vuik2019reproducing,pan2021quantized,cayao2021confinement} or unexpected properties missed by the analysis on pristine SM-SC nanowires~\cite{hui2015bulk,cole2016proximity,liu2018impurity,fleckenstein2018decaying,cao2019decays,kiendl2019proximity}. In real devices, disorder may arise from the surface oxidation of the SC, the imperfect substrates and gates, and more seriously, from the unintentional charged impurities~\cite{pantelides1978the}, the randomly distributed twin defects~\cite{caroff2009controlled}, and the stacking faults~\cite{shtrikman2009method} in SM nanowires. Typically, charged impurities in InAs and InSb nanowires could induce strong chemical potential fluctuations to the order of a few meV~\cite{woods2021charge}, which are much larger than the induced superconducting gap (around 0.2 meV), making it impossible to access the topological superconducting phase.

\begin{table}[t!]
\centering
\caption{Typical values of induced superconducting gap $\Delta_{\textrm{ind}}$, critical temperature $T_c$, and critical parallel magnetic field $B_{c,\parallel}$ extracted from tunneling spectroscopy measurements of hybrid SM-SC nanowires comprising InAs (or InSb) and different SC materials Al, Pb, and Sn.}\label{scpara}
\begin{tabular}{cccc}
\hline
\hline
~~~~~~~ &~~Al~~\cite{deng2016majorana,vaitiekenas2018effective,de2018electric}~~~~~ &~~Sn~\cite{pendharkar2021parity}~~~ &~~~Pb~\cite{kanne2021epitaxial}~~\\
\hline
$\Delta_\textrm{ind}$ (meV) &0.22---0.27 &0.7 &1.25 \\
\hline
$T_c$ (K) &$\sim 1$  &$\sim 3.7$ &$\sim 7$ \\
\hline
$B_{c,\parallel}$ (T) &$\sim 2$ &$\sim 4$ &$\sim 8.5$\\
\hline
\hline
\end{tabular}
\end{table}

Most recently, the electrostatic and electronic properties of PbTe, a IV-VI SM material, have been numerically studied~\cite{cao2022numerical} as a possible remedy for the interruption from disorders.
In comparison to InAs and InSb, PbTe has currently displayed several advantages.
To start with, PbTe is known to have a much larger dielectric constant ($\sim$ 1350) in comparison to that ($\sim$ 15) of InAs and InSb, leading to a higher tolerance of charged impurity in PbTe.
Meanwhile, as the sub-band of each valley can potentially host MZMs, the multiple-valley structure of PbTe provides more options (in comparison to the InAs and InSb situations) in the realization of the topological superconducting phases.

Additional advantages of PbTe-Pb over the existing hybrid nanowires can be expected: (i) Owing to the huge dielectric constant of PbTe, it has very high electron mobilities (about $10^6~\textrm{cm}^2/\textrm{Vs}$)~\cite{springholz1993mbe,ueta1997improved}, such that in submicron constrictions lithographically patterned in PbTe quantum wells, sequential quantized conductance steps were observed despite of a significant concentration of charged defects near the constrictions~\cite{grabecki1999quantum,grabecki2005disorder,grabecki2004ballistic,grabecki2006pbte}. (ii) A high-resolution transmission electron microscopy revealed that the epitaxially grown PbTe nanowires are free of stacking fault \cite{dziawa2010defect}. (iii) By virtue of the sizable parent gap $\Delta$ of Pb, a superconducting gap comparable with those of InAs-Al and InSb-Al can be induced in PbTe-Pb even by a weak coupling between Pb and PbTe, where the effective Land\'{e} g factor and spin-orbit coupling in PbTe do not get reduced much by the renormalization effects on SM from SC~\cite{stanescu2011majorana,cole2015effects,stanescu2017proximity,reeg2018metallization}. Meanwhile, at weak SM-SC couplings, the induced superconductivity in the SM is immune to any nonmagnetic disorder in the SC~\cite{hui2015bulk,cole2016proximity,liu2018impurity}. Finally, (iv) Pb shell on PbTe nanowire has less lattice mismatch than that of InAs-Al and InSb-Al. The lattice matching can even be further improved by buffering a thin CdTe layer between PbTe and Pb, as CdTe and PbTe share almost the same lattice constant.
The advantages mentioned above have inspired preliminary experimental efforts on the epitaxial growth and transport characterization of PbTe nanowires~\cite{schlatmann2021josephson,schellingerhout2022growth,geng2022observation,jiang2022selective} and hybrid PbTe-Pb nanowires~\cite{jiang2022selective}.

\begin{figure*}[t!]
\centering
\includegraphics[width=2\columnwidth]{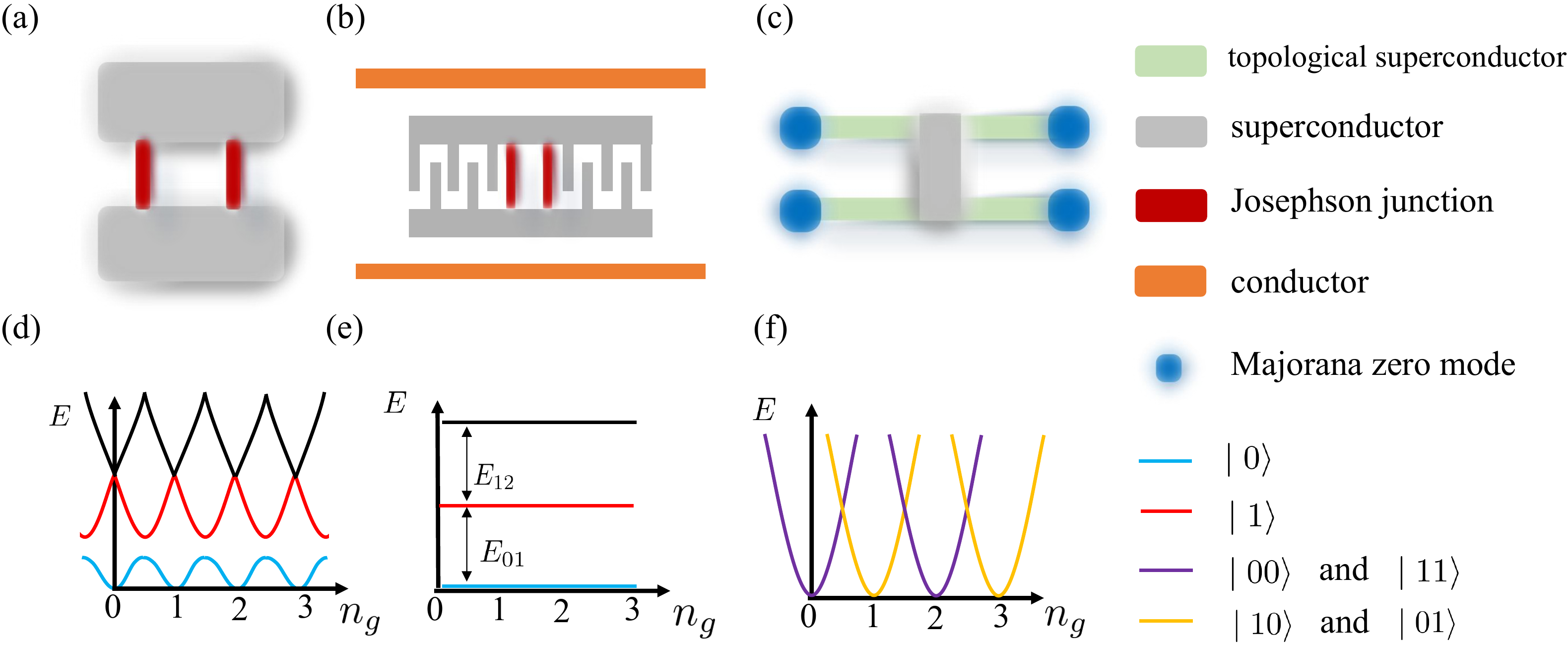}
\caption{Schematic and energy curves (as a function of the effective dimensionless gate voltage $n_g$) of a charge qubit [(a) and (d)], transmon qubit [(b) and (e)], and topological qubit [(c) and (f)]. Qubits of the first two types are based on conventional SCs while the third ones require topological SCs. States $\vert 0\rangle$ and $\vert 1\rangle$ [in (d) and (e)] of both the charge and transmon qubits refer to the system ground state and the first excited state, respectively.
In a charge qubit (d), both state energies and the energy difference $\Delta E$ between the ground state and the first excited state oscillate with $n_g$.
The oscillation of energies increases the complexity of the qubit manipulation. Indeed, when $n_g$ is tuned to around the odd values of (d), $\Delta E$ becomes much larger than the energy difference between the first and second excited states, making it hard to control the qubit state.
For transmon qubit in (e), the problem with oscillating energies is absent.
However, in the transmon qubit setup, the energy difference $ E_{12}$ is very close to $ E_{01}$. As a consequence, state leakage may occur when controlling the qubit with fast pulses. A topological qubit (f) contains four MZMs that combine into four degenerate states.
Two of them ($\vert 00\rangle$ and $\vert 11\rangle$) are in the even parity, and the other two ($\vert 01\rangle$ and $\vert 10\rangle$) are in the odd parity.
States with the same parity (either even or odd) have the same energy that is different from that with the opposite parity [see curves of (f)].
Qubit manipulation is realized between degenerate states with the same parity.
From (f), it can be seen that since the qubit is encoded within the same parity, charge noise then does not influence the qubit.
}\label{Fig:qubits}
\end{figure*}

\subsection{Material growth}
As one recent progress in the hybrid SM-SC nanowire fabrication, 
the integration of the selective-area growth technique with the shadow deposition of SC~\cite{jung2021universal,jiang2022selective} improves the quality of fabricated nanowire networks.
Previously, with the out-of-plane growth technique, nanowire networks were constructed by merging multiple wires during the growth~\cite{fadaly2017observation}.
To guarantee the phase-coherent and ballistic transport of the network, the out-of-plane growth technique has an extremely high requirement (with accuracy to the order of nanometer) on the positioning of the grown nanowires.
The requirement of high precision of the out-of-plane growth technique limits its application in the design of more complex networks.
The requirement of accuracy is however relaxed by the in-plane selective area growth technique~\cite{krizek2018field,aseev2018selectivity,op2020plane}.
With this technique, an amorphous mask, on which the trenches
for SM network growth can be opened, is firstly deposited on the substrate.
The right growth conditions confine the subsequent nanowire synthesis to the mask openings, allowing for scalable and highly flexible growth of complex in-plane selective area networks. Subsequently, a SC shell was epitaxially grown on the surface of the SM nanowire networks and then selectively etched for covering metallic contacts or the tuning of electrostatics by voltage gates. The etching process generally damages the SM surface and thus prevents ballistic electron transport. To overcome this shortcoming, several shadow walls, with designed heights and widths can be fabricated before the SM nanowire network growth~\cite{jung2021universal,heedt2021shadow,jiang2022selective}.
The walls act as shadowing objects during the subsequent SC deposition and allow for selective growth on the SM network.

\section{Topological quantum computation based on hybrid SM-SC nanowires}\label{Sec V}
Qubit is the fundamental building block in quantum computation. In theory, a qubit can be constructed by any two-state system with states denoted as $\vert 0\rangle$ and $\vert 1\rangle$. These states can be, e.g., the occupied and unoccupied status of an energy level, the spin-up and spin-down of an electron, or the left and right polarization of a photon. One of the most formidable difficulties for almost all existing qubit platforms is the decoherence problem caused by inevitable environmental noises~\cite{nielsen2002quantum}. By contrast, MZMs can form topological qubits that store information nonlocally. Specifically, $N$ pairs of spatially well-separated MZMs in a topological material will form a $2^{N-1}-$ fold degenerate ground state subspace, which is separated from the high-energy subspace by a topological gap. If the temperature is much smaller than the topological gap, the information encoded in the degenerate ground state subspace becomes insensitive to quantum decoherence induced by local noise or fluctuations: this is the so-call topological protection~\cite{kitaev2003fault,nayak2008non,sarma2015majorana}. In addition, due to the non-Abelian exchange statistics obeyed by MZMs~\cite{moore1991nonabelions,nayak19962n}, braiding of MZMs along with the topological charge measurement (e.g., fermion parity measurement in topological SCs) can generate complete fault-tolerant Clifford gate operations~\cite{kitaev2003fault,nayak2008non}. Certain non-Clifford gates, e.g., T-gate, can be realized by preparing noisy magic states along with state distillation protocols~\cite{Bravyi&Kitaev-Magic}. Consequently, Majorana qubit system becomes an important candidate to realize a universal fault-tolerant quantum computation. In this section, we show the advantages of topological qubits over charge qubits~\cite{Bouchiat_1998-CPB} and transmon qubits~\cite{Koch-transmon}, the latter two can be constructed by conventional $s$-wave SCs.

Two adjacent SCs connected by a normal conductor can form a so-called JJ. As shown in Fig.~\ref{Fig:qubits}(a), two parallel JJs made of mesoscopic superconducting islands can be considered as a charge qubit~\cite{jacobs2014quantum}.
Two lowest-energy eigenstates of the system play the role of the qubit states $\vert 0\rangle$ and $\vert 1\rangle$.
In the charge qubit, there exist two relevant competing energy scales: the Josephson coupling $E_J$ across the junction and the Coulomb charging energy $E_C$. For $E_J=E_C$, as shown in Fig.~\ref{Fig:qubits}(d), the low-lying energies strongly depend on the gate voltage $V_g$ parameterized with a dimensionless quantity $n_g=CV_g/e$, where $C$ is the capacitance of the island. This remarkable charge dispersion amounts to that the charge qubit is sensitive to charge noise~\cite{ChargeNoise}. Theoretically, the charge dispersion can be reduced exponentially by increasing the ratio $E_J/E_C$. It was proposed that enhanced $E_J/E_C$ is accessible in a special design depicted in Fig.~\ref{Fig:qubits}(b)~\cite{Koch-transmon}, in which a charge qubit created by sawtoothed superconducting islands is shunted by two transmission lines. This setup is called transmon qubit~\cite{Koch-transmon}. As shown in Fig.~\ref{Fig:qubits}(e), when $E_J$ is an order larger than $E_C$, the charge dispersion is almost completely eliminated. However, in the transmon qubit setup, the energy difference between the second and the first excited states ($ E_{12}$) is very close to that between the first excited and the ground states ($ E_{01}$), similar to the spectrum of a 1D harmonic oscillator. When the transmon qubit is controlled by fast pulses with wide frequency response, this low anharmonicity of transmon qubit may lead to leakage out of the computational subspace defined by the two lowest-lying energy eigenstates~\cite{steffen2003accurate,motzoi2009simple}. In addition, although decoherence caused by charge noise can be resolved, there are other types of decoherence in transmon qubit, e.g., due to quasiparticle poisoning~\cite{Catelani-QP-transmon,QP-transmon-Exp}. We note that hybrid SM-SC nanowires have been employed to create gate tunable transmon qubit, known as gatemon qubit~\cite{larsen2015semiconductor,de2015realization,luthi2018evolution,sabonis2020destructive,van2020photon} and see Ref.~\cite{aguado2020perspective} for a review. The combination between transmon qubits and Majorana-based qubits has been proposed~\cite{Hassler_2011,van_Heck_2012,HyartPRB2013,ginossar2014microwave,yavilberg2015fermion,keselman2019spectral,avila2020majorana,avila2020superconducting} to construct topological transmon qubits with topological protection against local sources of noise.
In a Majorana box qubit (MBQ)~\cite{plugge2017majorana}, also named as two-sided tetron~\cite{karzig2017scalable}, consisting of four MZMs at the ends of two parallel topological SCs shown in Fig.~\ref{Fig:qubits}(c), both the topological degeneracy~\cite{fu2010electron} shown in Fig.~\ref{Fig:qubits}(f) and the charging energy can resolve both the decoherence and quasiparticle poisoning problems at once.

\begin{figure}[t!]
\includegraphics[width=0.9\columnwidth]{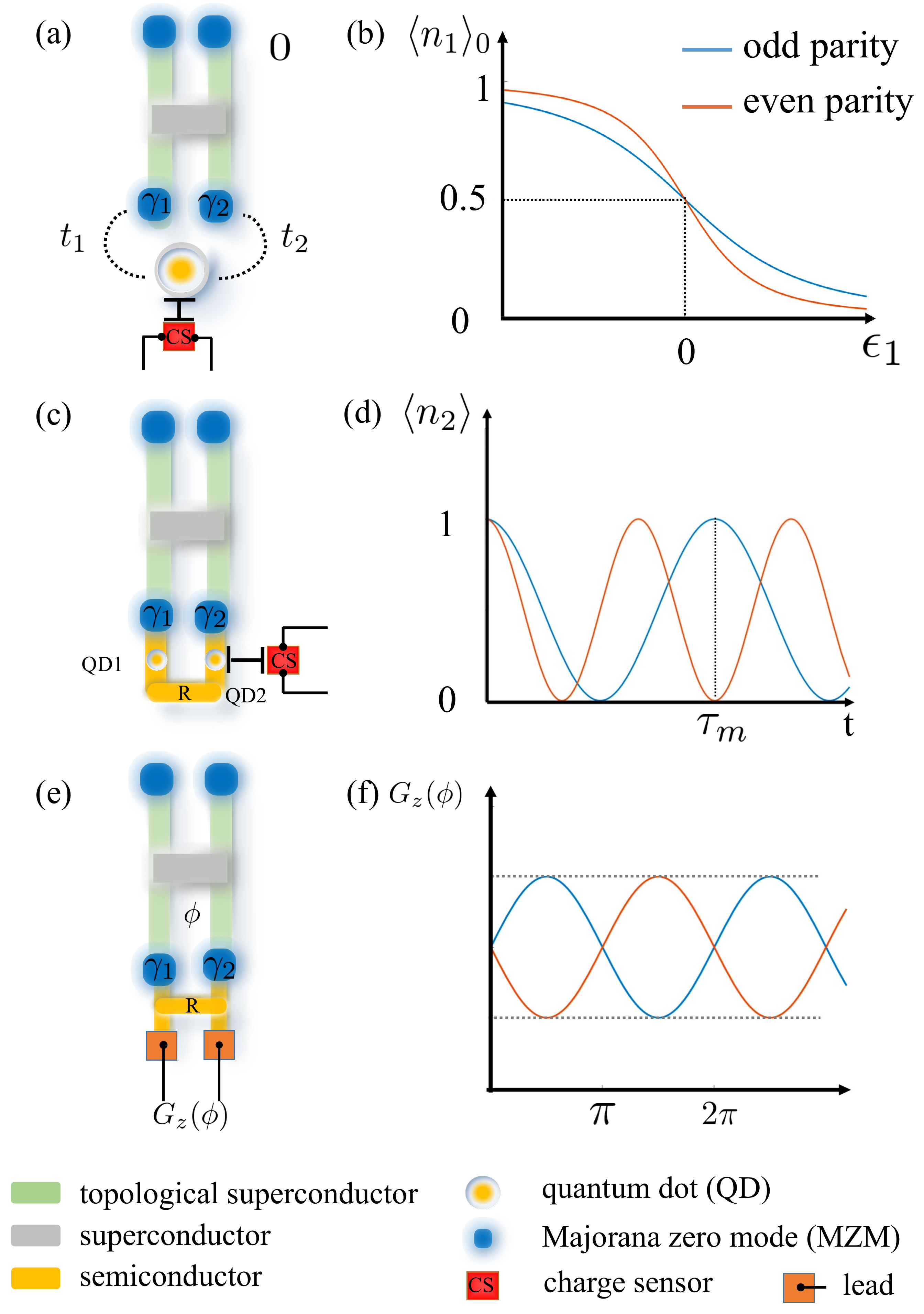}
\caption{Three Majorana parity readout protocols and their corresponding
parity-dependent curves.
(a) A protocol to read out the Majorana parity by measuring the occupation (via the charge sensor labeled by ``CS'') of the quantum dot that couples to two MZMs. The dot has the energy $\epsilon_1$.
(b) For MZMs with different parities, the dot occupation $\langle n_1 \rangle_0$ follows different $\epsilon_1$-dependent curves. One can detect the MZM parity as long as the dot is off-resonance, i.e., $\epsilon_1 \neq 0$.
(c) The protocol that detects the MZM parity by coupling the charge sensor to a quantum dot (QD2).
(d) After tuning QD1 and QD2 to be degenerate at time $t = 0$, the dot state Rabi-oscillates between two states.
The time $\tau_m$ of a projective measurement indirectly conveys the information of the MZM parity.
(e) Majorana parity readout from interference between two trajectories: the one through the reference arm R (yellow) and that through the SC (gray).
(f) The corresponding conductance curves $G_{z}(\phi)$ as functions of the dimensionless magnetic flux.
}\label{Fig:protocol}
\end{figure}

As mentioned previously, MZMs can potentially provide a solution for realizing fault-tolerant quantum computation. MBQ scheme~\cite{plugge2017majorana,karzig2017scalable} is one of the most promising schemes among various MZM-based topological qubit proposals. The qubits in this scheme are actually topologically protected superconducting charge qubits, which include two or more parallel hybrid SM-SC nanowires bridged by conventional SCs.
One can imagine this MBQ device as a mesoscopic topological superconducting island. This island has a constant Coulomb charging energy and is connected to other constituents, i.e., trivial SCs and other MBQs, via JJs.
Most interestingly, when the charging energy is much greater than the Josephson energy,
the system is in a topological Cooper pair box regime~\cite{plugge2017majorana,karzig2017scalable}, where the nonequilibrium quasiparticle tunneling (or quasiparticle poisoning) outside the MBQ can be significantly reduced.
This MBQ scheme is thought to be better than the earlier topological transmon qubit schemes~\cite{Hassler_2011,van_Heck_2012,HyartPRB2013,ginossar2014microwave}, which are more easily affected by quasiparticle poisoning outside the qubit island~\cite{Catelani-QP-transmon,QP-transmon-Exp}.
Usually, each MBQ contains four MZMs (or six MZMs, where two extra MZMs are ancillary ones to realize the Majorana braiding protocol~\cite{ClarkeBraiding}) located at the wire ends. These MZMs are described by Majorana fermion operators $\gamma_j=\gamma_j^\dag$, where $j=1,2,3,4$.
Due to the strong charging energy on the island, the fermion parity of the island is a conserved quantity. Supposing that the four MZMs are spatially well separated and the parity is fixed as even or odd, the ground state of the MBQ is two-fold degenerate and behaves as a spin-1/2 freedom. Consequently, due to the non-Abelian exchange statics obeyed by MZMs~\cite{moore1991nonabelions,nayak19962n}, Pauli operators for the MBQ can be represented by Majorana fermion operators~\cite{plugge2017majorana}: $\sigma_x=i\gamma_1\gamma_2$, $\sigma_y=i\gamma_3\gamma_1$, and $\sigma_z=i\gamma_2\gamma_3$. 

\begin{figure}[t!]
\includegraphics[width=\columnwidth]{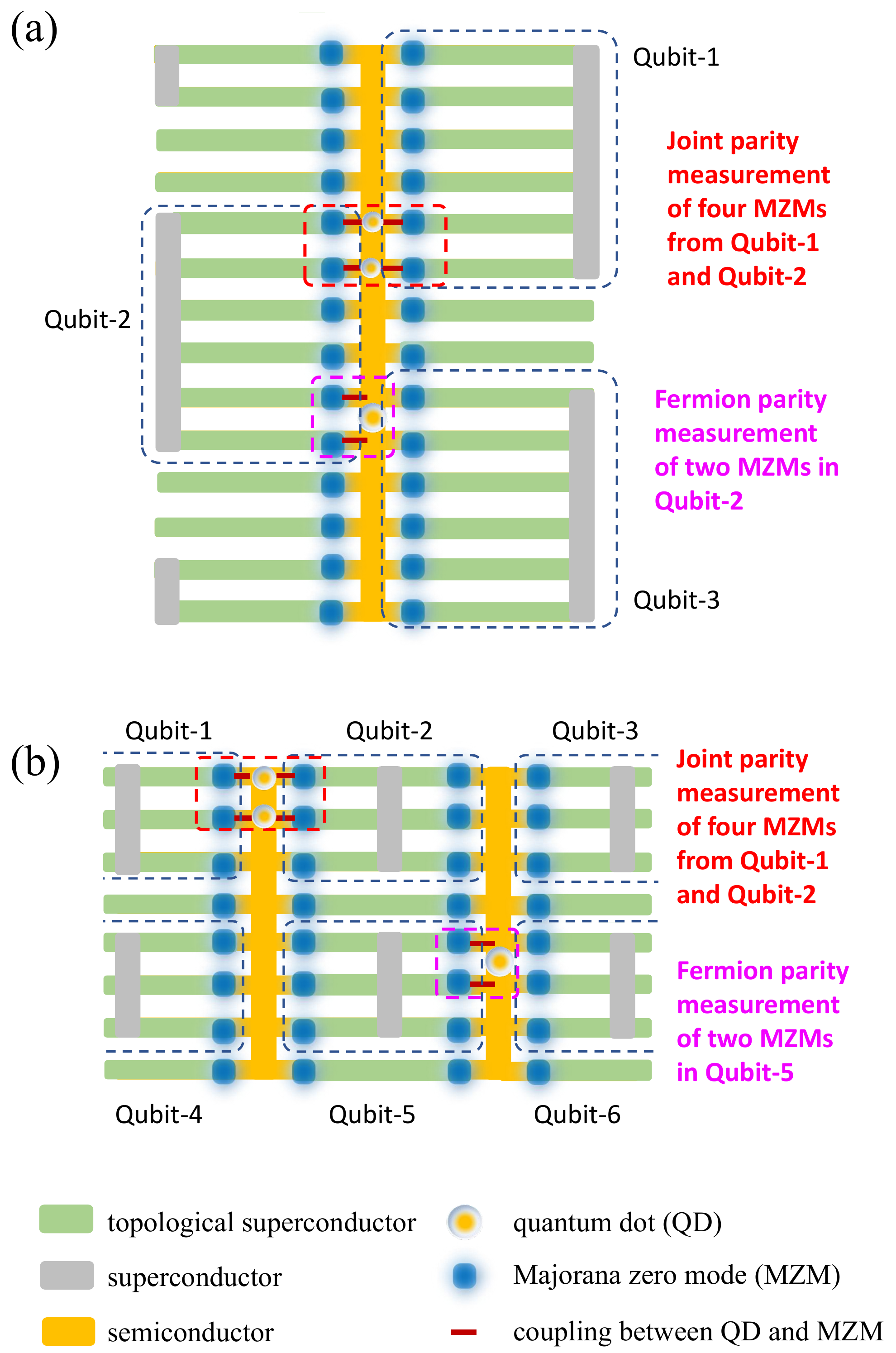}
\caption{Scalable designs of topological quantum computation networks~\cite{karzig2017scalable} based on the protocol of Fig.\,\ref{Fig:protocol}(a). With the dot-MZM couplings switched on, the Fermion parity of two MZMs and the joint parity of four MZMs can be obtained from either the charge sensing or the differential capacitance measurement of the quantum dot (not shown in this figure).
(a) The one-sided hexon architecture.  
(b) The two-sided hexon architecture. For both cases, each qubit (in each dashed box) consists of six MZMs connected by a conventional SC (gray).
}\label{Fig:scalable}
\end{figure}

In quantum computation, qubit readout represents a critical issue. Below, we briefly review three major protocols of MBQ readout, while a few other readout protocols were reviewed in Ref.~\cite{beenakker2020search}.
The first MBQ readout protocol~\cite{karzig2017scalable}, as shown in Fig.~\ref{Fig:protocol}(a), includes a quantum dot that tunnel-couples to two MZMs $\gamma_1$ and $\gamma_2$ of the MBQ (with amplitudes $t_1$ and $t_2$, respectively), and also weakly couples to a charge sensor.
When the gate voltage $n_g$ of the dot (not shown) is tuned to the regimes where the dot-MBQ system is near on-resonance, the average charge on the dot shows different $n_g$-dependence, for states with different parities [Fig.~\ref{Fig:protocol}(b)]. Thus, the detection of the dot occupation by a nearby charge sensor at different gate voltages enables the readout of the MBQ parity.
In addition, the qubit readout can also be realized in this setup by measuring the dot differential capacitance~\cite{karzig2017scalable}. The setup of the second MBQ readout protocol~\cite{plugge2017majorana} is shown in Fig.~\ref{Fig:protocol}(c), in which two quantum dots are tunnel coupled to their nearest MZMs $\gamma_1$ and $\gamma_2$ in the MBQ. Two dots are also inter-coupled through a reference arm $R$. Physically, electrons can tunnel back and forth between the two dots either through the box or via the arm. As a result, the average charge on dot 2 oscillates in time with a period that depends on the MBQ parity, as shown in Fig.~\ref{Fig:protocol}(d). A charge sensor near dot 2 can perform a projective measurement on the instantaneous average charge of dot 2, to obtain the MBQ parity. We mention that at some special moments, e.g., $t=\tau_m$, where the extremes of the curves in Fig.~\ref{Fig:protocol}(d) arrive, the qubit readout is ideal. Alternatively, the readout in frequency can be achieved with a resonator that capacitively couples to dot 2~\cite{plugge2017majorana}. Interestingly, in a similar two-dot setup, the qubit information can be inferred from the charge states of two dots through a Landau-Zener transition~\cite{LZ-MZMreadout}.
In this case, the charge sensing readout is nondemolition and robust against the low-frequency fluctuations~\cite{LZ-MZMreadout}.
The third MBQ readout protocol~\cite{plugge2017majorana} can be implemented in a dot-free setup under a tunable magnetic flux $\phi$, as shown in Fig.~\ref{Fig:protocol}(e).
With a small bias voltage applied between the two terminals of such an Aharonov–Bohm ring island, the interplay between electron cotunneling and interference dominates the transport. As shown in Fig.~\ref{Fig:protocol}(f), the measured $\phi$-dependent conductance can clearly reveal the even and odd MBQ parity. 

In practice, the potential weakness of each readout protocol above may pose challenge for experimental implementation. For the third protocol, phase coherence in the reference arm $R$ requires small bias voltage, such that cotunneling conductances are small, thus this current-based readout schemes would be limited by the time needed for data accumulation~\cite{plugge2017majorana}. For the second protocol, this single-shot projective measurement of the charge of the quantum dot is ideal when the occupation or inoccupation probability vanishes at particular times, i.e., $\tau=\tau_m$ in Fig.~\ref{Fig:protocol}(d), thus it requires fine tuning and rapid manipulation of system parameters~\cite{plugge2017majorana}. Simply re-measure multiple times can not improve the readout fidelity~\cite{plugge2017majorana}. For the first protocol, the charge state of the quantum dot is only weakly correlated with the state of the MBQ thus the readout requires longer measurement time~\cite{steiner2020readout}. 
In addition, one has to repeatedly couple and decouple the MBQ and quantum dot, which is 
also prone to errors. However, a recent theoretical work proposed that a projective readout of the MBQ can indeed be robustly implemented even though the MBQ is coupled to the quantum dot during the entire readout procedure~\cite{steiner2020readout}. Recently, the effects of noise~\cite{khindanov2021visibility} and Markovian dynamics~\cite{munk2020parity} on the measurement of this readout protocol have been studied. There exist crucial remarks on the feasibility of the first readout protocol: Charge measurement is very fast and accurate, and represents a well-understood technique in the SM community and is compatible with large magnetic fields~\cite{karzig2017scalable}. Additionally, the inclusion of charge sensors in the qubit plane does not prevent from scaling the system to a two-dimensional array of qubits~\cite{karzig2017scalable}.

With the first readout protocol introduced above, one can detect the parity of a single qubit, and the joint parity of two neighbouring qubits with designed scalable structures~\cite{karzig2017scalable}.
Several scalable designs with the Majorana qubits are proposed in Refs.~\cite{karzig2017scalable,Vijay&Fu-Majorana}, including the one-sided and two-sided hexon structures (see Fig.\,\ref{Fig:scalable}), and
the linear and two-sided tetron structures (not shown).
In the hexon architectures, one computational qubit consists of four MZMs and two ancillary ones. The latter two MZMs are included to assist Majorana braiding~\cite{ClarkeBraiding}.
Figures \ref{Fig:scalable}(a) and \ref{Fig:scalable}(b) present the one-sided hexon and the two-sided hexon architectures~\cite{karzig2017scalable}.
If the couplings between a quantum dot and two MZMs are switched on, the parity of the corresponding two MZMs can be read out from the quantum dot measurement as shown in previous paragraphs.
The joint parity of four MZMs can be obtained using two quantum dots, where the couplings between the dot and two neighboring MZMs are switched on for both dots [see Figs.\,\ref{Fig:scalable}(a) and \ref{Fig:scalable}(b)]. References \cite{karzig2017scalable,Vijay&Fu-Majorana} also demonstrate that braiding operations can be realized by applying a sequence of Majorana parity measurements.
In theory, all single-qubit and two-qubit Clifford operations are then realizable in hexon architectures using two-MZM parity measurements and four-MZM joint parity measurements, without the need of performing braidings in real space. 

\section{Summary and perspectives}\label{Sec VI}
To summarize, we have reviewed the recent progress on (i) the engineering and detection of MZMs in SM-SC heterostructures, and (ii) the MBQ-based topological quantum computation. Briefly, after an extensive effort in the last decade, the theory on the MZM realization has been widely accepted; several challenges however remain on the experimental side, especially due to the extremely high sample requirements. Among the reviewed SM-SC heterostructures, the most promising one is perhaps the partially covered SM-SC nanowire, due to its simple realization and easy tunability. It also has a closer connection to our ultimate purpose: the realization of topological quantum computation, following the architectures shown in Fig.~\ref{Fig:scalable}.

On the MZM detection, conclusive experimental evidence of MZMs in the SM-SC heterostructures remains missing. The major difficulty in the MZM detection comes from two reasons. The first reason is that the topological phase and MZM are significantly influenced by the control capability of the electrostatic potential~\cite{mikkelsen2018hybridization,woods2018effective,antipov2018effects} and the disorder effects~\cite{sarma2021disorder}. In fact, a low charged impurity density or a high tolerance to charged impurity is required for a hybrid SM-SC device to enter the topological regime.
This requirement might be satisfied by choosing new
SM and SC materials. For instance, Refs.~\cite{cao2022numerical,jiang2022selective,schlatmann2021josephson} has investigated the possibility to engineer MZMs in a hybrid PbTe-Pb nanowire.
As for the fabrication of the complex nanowire network shown in Fig.~\ref{Fig:scalable}, it is within reach of the selective area growth technique~\cite{krizek2018field,aseev2018selectivity,op2020plane,jiang2022selective,geng2022observation} that employs a pre-designed mask to define a network with an arbitrary pattern. The second reason is that Majorana signatures, e.g. the
%simplest 
most well-known zero bias conductance in the standard tunneling spectroscopy,
are easily contaminated or buried by false positive signatures due to ABSs or other disorder effects. Those false positive signatures in the detection can be resolved by introducing a dissipative environment in the tunneling~\cite{liu2013proposed,liu2022universal,zhang2022supressing}, where the non-Majorana signals will be suppressed and the Majorana signals remain. In addition, the combination of local and nonlocal measurements in the same device, e.g. the recent Microsoft experiment~\cite{aghaee2022inas}, could significantly increase the reliability of the experimental data. Finally, we want to emphasize that any experiments, including those listed in Sec.~\ref{Sec III}, can greatly aid the development of this area, once data with better precision can be achieved.

\section{Acknowledgements}
This work was supported by the National Natural Science Foundation of China (Grants No.~12004040 and No.~11974198), the Innovation Program for Quantum Sci-
ence and Technology (Grant No. 2021ZD0302400), and Tsinghua University Initiative Scientific Research Program.

\bibliographystyle{apsrev4-1}
\bibliography{refs-Majorana}

\end{document}